# Modal approach for tailoring the absorption in a photonic crystal membrane

Romain Peretti [1], Guillaume Gomard[1], Christian Seassal[1], Xavier Letartre[1], Emmanuel Drouard[1]*

*1-Université de Lyon, Institut des Nanotechnologies de Lyon (INL), UMR 5270, CNRS-INSA-ECL-UCBL, France Ecole Centrale de Lyon, 36 Avenue Guy de Collongue, 69134 Ecully Cedex, France*

In this paper, we propose a method for tailoring the absorption in a photonic crystal membrane. For that purpose, we first applied Time Domain Coupled Mode Theory to such a subwavelength membrane and demonstrated that 100% resonant absorption can be reached even for a symmetric membrane, if degenerate modes are involved. Design rules were then derived from this model in order to tune the absorption. Subsequently, Finite Difference Time Domain simulations were used as a proof of concept and carried out on a low absorbing material (extinction coefficient=$10^{-2}$) with a high refractive index corresponding to the optical indices of amorphous silicon at around 720 nm. In doing so, 85% resonant absorption was obtained, which is significantly higher than the commonly reported 50% maximum value. Those results were finally analyzed and confronted to theory so as to extend our method to other materials, configurations and applications.

# I. INTRODUCTION

Controlled light absorption is a topic of particular importance for applications such as photovoltaic solar [1] and indoor cells [2] or sensors [3]. In order to decrease the costs and facilitate the integration of those devices, the use of thin layers of active materials is preferred. Unfortunately, the single pass absorption of light in such layers is inefficient, especially at wavelengths close to the material bandgap, where the extinction coefficient (denoted κ) can be low, whereas the reflectance may remain high due to a large index contrast with the surrounding medium.

In this case, the use of resonant optical modes is the key to enhance the absorption [1], since it can simultaneously increase the light path in the absorbing layer and reduce its reflectance. Such modes can typically be introduced into these layers thanks to corrugations such as Photonic Crystals (PhC), which are known to allow a coupling between the incident light and some “guided mode resonances” [4]. In the following, these structured layers will be called “photonic membranes”.

Thus, a usual phenomenological method to analyze the effect of resonant optical modes is the Time Domain Coupled Mode Theory (TDCMT) [5, 6] already used for an extensive analysis of different photonic structures [7, 8]. It was first introduced in [9] and applied to add-drop filters [5] as well as absorbing systems [6, 10, 11].

Using this method, it has been shown that the absorption in a single, isolated and symmetric photonic membrane is limited to 50%. Higher (up to 100%) absorptions were also predicted, but for membranes integrating a back reflector [12, 6].

However, it appears that the 50% absorption limit can be exceeded even for a symmetric absorbing membrane using the contribution of several modes [11], as will be discussed in the following. In the second section of the paper, we derive the TDCMT for a single membrane and show how this limit can be overcome even with such a simple system. We highlight the role played by the coupling anisotropy and demonstrate that two degenerate modes can lead to an absorption as large as 100% for a given wavelength. In the third section, we illustrate this concept using a 2D PhC membrane, a structure known to enhance absorption [13] and that can be easily fabricated by various methods including holographic [14], e-beam [15] or nano-imprint [16] lithography. Finite Difference Time

Domain (FDTD) simulations are performed on this patterned membrane so as to understand the influence of its main parameters on the modal properties as shown on section IV. A methodology is then derived out of those simulations in order to taylor the resonances taking into account the material constraints. In low absorbing materials, the optical indices determine the modal properties needed to optimize the absorption. Those modal properties can be tuned by modifying the PhC membrane parameters (its filling factor, thickness and period). In doing so, general design rules could be provided and an absorption of 85% was obtained for the optimized structure.

As highlighted in the last section of the paper, those simulation results are completely consistent with the TDCMT and can be directly used to predict the absorption of a thin membrane with only a limited set of data.

# II. TIME DOMAIN COUPLED MODE THEORY DESCRIPTION OF AN ABSORBING MEMBRANE

In this section, the absorption of a photonic membrane is derived thanks to the TDCMT. The photonic membrane is defined here as a photonic resonator exhibiting resonant modes which can be coupled to only one propagative mode. It can be, for instance, a PhC membrane vertically addressed through its Bloch modes at the Γ point of the first Brillouin zone (corresponding to a null in-plane wave vector). In this paper, PhC membranes are supposed to be etched over the whole thickness, with vertical walls. They are thus symmetric with respect to their midplane. Owing this symmetry, such a photonic membrane can absorb at most 50% [10, 6, 11], if only one resonant mode is involved. This limit can however be overcome to reach theoretically a 100% total absorption, by combining two orthogonal modes.

Following the formalism developed in [5], we assume that the resonant modes of any photonic membrane can couple with the propagative free space waves of amplitudes $S_{+1}$ (incident), $S_{-1}$ (reflected) $S_{-2}$ (transmitted) as shown in FIGURE 1.

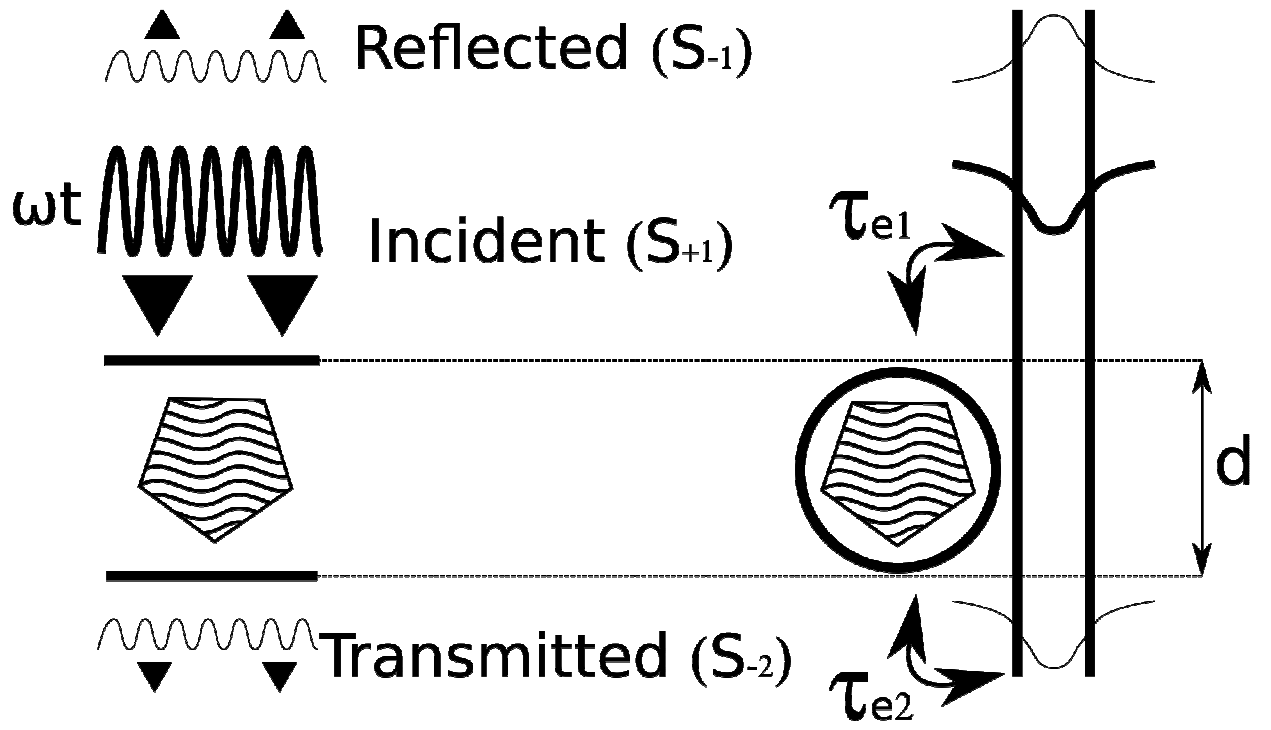

**FIGURE 1 : LEFT: SCHEMATIC VIEW OF THE ABSORBING THIN FILM RESONATOR STUDIED IN THE FRAME OF THE TDCMT. RIGHT: SAME MODEL TRANSLATED IN TERMS OF MICRORESONATORS COUPLED TO A WAVEGUIDE AS DEPICTED IN [5]. NO SPECIFIC ASSUMPTION IS MADE ABOUT THE SYMMETRY OF THE MODE.**

## Single mode membrane

A single mode resonator is introduced without any assumption on its symmetry properties. As in [5], and using previous notations, the equations describing its temporal evolution can be written as:

$$\frac{da}{dt} = \left( j\omega_0 - \frac{1}{\tau_0} - \frac{1}{\tau_e} \right) a + \kappa_1 S_{+1}$$
$$S_{-1} = -e^{-j\beta d} \kappa_2^* a \tag{1}$$
$$S_{-2} = e^{-j\beta d} \left( S_{+1} - \kappa_1^* a \right)$$

where $\omega_0$ is the mode frequency, $\tau_0$ the decay rate due to absorption losses, $\tau_e$ the decay rate due to the coupling with reflected and transmitted modes (in the following we defined $\tau_{e_1}$ and $\tau_{e_2}$ as " the decay rates in the forward (frontside) and backward (backside) direction" as in [5]. Those parameters are linked through the relation $2\tau_e^{-1} = \tau_{e_1}^{-1} + \tau_{e_2}^{-1}$). $\kappa_1$ ( $\kappa_2$ ) is the coupling coefficient with the forward (respectively backward) plane wave that is proportional to $\tau_{e_1}^{-1/2}$ (respectively $\tau_{e_2}^{-1/2}$ ), $\beta$ is the propagation constant of the wave along the membrane cross-section, and $d$ the membrane thickness.

Stating $(\omega - \omega_0) = \delta\omega$, one can express the reflection, transmission coefficients in energy and the resulting absorption in the membrane as:

$$R = \frac{1}{\tau_{e_2}\tau_{e_1}} \left( \left( \frac{1}{\tau_0} + \frac{1}{\tau_e} \right)^2 + \delta\omega^2 \right)^{-1}$$
$$T = 1 + \frac{1}{\tau_{e_1}} \frac{\frac{1}{\tau_{e_1}} - 2\left( \frac{1}{\tau_0} + \frac{1}{\tau_e} \right)}{\left( \frac{1}{\tau_0} + \frac{1}{\tau_e} \right)^2 + \delta\omega^2} \tag{2}$$
$$A = 1 - R - T = 2\left[ \left( \left( \frac{1}{\tau_0} + \frac{1}{\tau_e} \right)^2 + \delta\omega^2 \right) \tau_{e_1} \tau_0 \right]^{-1}$$

A maximum of absorption is obtained provided two conditions are met: To work (1) at the resonant wavelength of the mode ( $\delta\omega = 0$ ), and, as can shown by derivation (2) at the critical coupling conditions defined by $\tau_0 = \tau_e$ [11]. This gives:

$$A_{\max} = \left[ 1 + \frac{\tau_{e_1}}{\tau_{e_2}} \right]^{-1} \tag{3}$$

In addition to the previous conditions, it can be seen that the maximum absorption is directly related to the anisotropy of the mode coupling with the propagative waves, i.e. the incident and reflected waves on one hand, and the transmitted wave on the other hand. In the case of a PhC membrane, the membrane is absolutely symmetric and non-degenerate modes can only be symmetric or anti-symmetric, leading to an only 50% maximum absorption.

Thus, the most important feature for a mode to overcome the usual 50% absorption limit is its coupling anisotropy which enables to reach total absorption. A way to achieve this is to introduce asymmetry in the system thanks to a back reflector [6], an anisotropic patterning [17, 18] or, as will be presented below, to use orthogonal degenerate modes.

## Two orthogonal degenerate modes membrane

Because most of the resonators support a collection of modes and not only a single one, it is relevant to study the effect of their interaction on the absorption properties. For the sake of simplicity, we restricted this study to the case of two orthogonal modes contributing to the absorption, one being vertically symmetric (denoted with subscript S) and the other one being anti-symmetric (subscript A as shown in Figure 2).

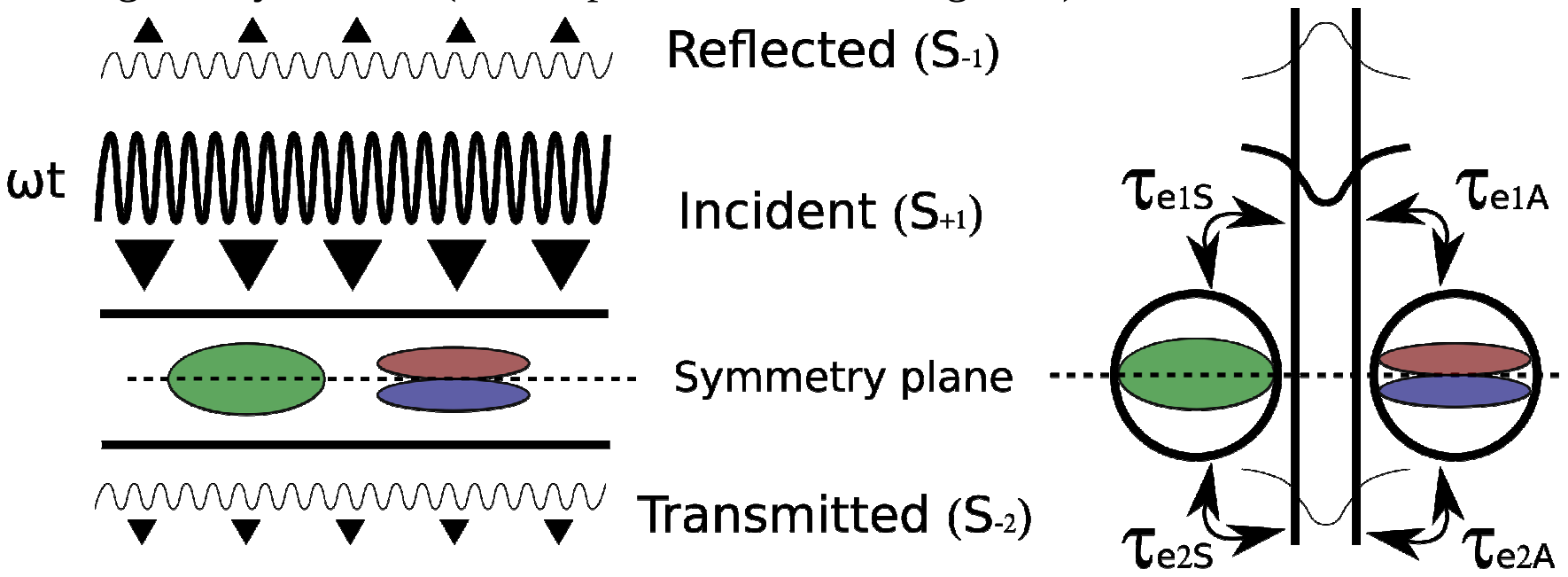

**FIGURE 2 : SIMILAR SCHEMATIC AS IN FIGURE 1 AFTER INTRODUCING BOTH A SYMMETRIC AND AN ANTI-SYMMETRIC MODE.**

Nevertheless, the following results remain valid for any couple of orthogonal modes. Because of the orthogonality, the direct coupling between those two modes is forbidden and the coupling equations system is now written:

$$\begin{aligned}
\frac{da_S}{dt} &= \left( j\omega_0 - \frac{1}{\tau_{0S}} - \frac{1}{\tau_{eS}} \right) a_S + \kappa_{1_S} S_{+1} \\
\frac{da_A}{dt} &= \left( j\omega_0 - \frac{1}{\tau_{0A}} - \frac{1}{\tau_{eA}} \right) a_A + \kappa_{1A} S_{+1} \\
S_{-1} &= e^{-j\beta d} \left( \kappa_A^* a_A - \kappa_S^* a_S \right) \\
S_{-2} &= e^{-j\beta d} \left( S_{+1} - \kappa_S^* a_S - \kappa_A^* a_A \right)
\end{aligned} \tag{4}$$

As previously, the field energy reflection, transmission coefficients and resulting absorption can be calculated and we find:

$$\begin{aligned}
R &= R_S + R_A - 2\Re\left[ \left( \frac{S_{-1S}}{S_{+1S}} \right) \left( \frac{S_{-1A}}{S_{+1A}} \right)^* \right] \\
T &= T_S + T_A - 1 + 2\Re\left[ \left( \frac{S_{-1S}}{S_{+1S}} \right) \left( \frac{S_{-1A}}{S_{+1A}} \right) \right] \\
A &= A_A + A_S
\end{aligned} \tag{5}$$

This result indicates that it is possible to achieve a 100% absorption by combining a symmetric and an anti-symmetric mode provided that they are degenerate and both in the critical coupling regime. It can also be pointed out that in order to derive the results above, only the orthogonality hypothesis was assumed. Therefore, those expressions are still valid for any couple of degenerate modes which are orthogonal, regardless of their symmetry.

To conclude this section, we first showed that the usually admitted 50% absorption limit can be overcome in a simple PhC absorbing thin membrane. We showed that an important parameter is the coupling anisotropy of the mode. Finally we showed that a way to introduce this coupling anisotropy is to use degenerate modes coming from one symmetric and from one anti-symmetric mode.

# III. PHC MEMBRANE MODES SELECTION USING FDTD SIMULATIONS

*Required properties of the PhC membrane modes for absorption enhancement*

A 2D PhC membrane, as shown on Figure 3 (left), is a good candidate to illustrate how one can reach a total absorption as predicted by the analytical model. Indeed, as shown in the following, the numerous modes it offers have adjustable quality factors, thus coupling coefficients and frequencies thanks to the tuning of their filling factor and thickness.

Since the membrane thickness is much lower than both the wavelength and the absorption length, the main absorption enhancement will originate from the *x*-*y*-plane propagation of the light inside the layer, and therefore from the Bloch modes of the PhC. The Bloch modes at the Γ point (which is commonly used for applications) of the dispersion diagram can be distinguished by several ways.

It is first noticeable that these modes are either symmetric (even) or anti-symmetric (odd) with respect to the midplane of the membrane. Then, some of the modes are purely guided modes and cannot couple to free space modes. In other words, their electric field is anti-symmetric in the *x*-*y*-plane [22]. The other modes have a non-zero order Bloch coefficient, which means that they exhibit a non-zero overlap with the incident electric field of the plane wave at the interfaces. This resulting overlap can even be tailored thanks to the geometric parameters of the PhC membrane, and so can be the coupling. The respective main influence of the filling factor and thickness, for a given period fixing the resonant wavelength, are detailed in the following. Thus, an appropriately chosen couple of modes can satisfy all the conditions for a maximized absorption under vertical incidence, i.e a critical coupling for each mode, the degeneracy and the orthogonality.

It has to be noticed that all the aforementioned modes can not follow exactly the TDCMT model. Indeed, the PhC modes are embedded in Fabry-Perot cavity modes, due to the reflection at the membrane interfaces [12]. However, provided that the quality factor of the Fabry Perot mode is low compared to the one of the PhC mode, this cavity effect can be neglected. Thus, as shown in the following, the absorption resulting from the modes in 2D PhC membrane made of an absorbing material can be described using the TDCMT model.

For a given quality factor, the confinement inside the material of the considered PhC membrane modes has also an influence on the absorption enhancement: the stronger it is, the greater can be the enhancement. In the lack of any specific quantum electrodynamics interaction (modes remains delocalized), a lower limit of the required quality factor for critical coupling can be established by considering a full confinement, as it would be using a plane wave in an homogenous absorbing medium. Indeed, for given refractive index *n* and extinction coefficient $\kappa$; it can be shown [19] that:

$$Q_0 \geq n / 2\kappa \qquad (6)$$

*Selection of two modes thanks to their symmetry properties*

The parameters of the structure considered for the numerical illustration are the following: the membrane has a refractive index *n*=4.06, which corresponds to the refractive index of amorphous silicon (a-Si) at 720 nm [19], and $\kappa$=0 (non-absorbing material, in a first step). The 2D PhC has a square lattice (polarization insensitive under normal incidence), with an air surface filling factor *ff*=0.15 $\left( ff = \pi \frac{r^2}{a^2} \right)$ and a thickness *d*= *a/2*, where *a* is the period of the PhC and *r* the radius of the holes.

FDTD [20] simulations were performed to get its band diagram by introducing in-plane (*xy*) periodic boundary conditions. The analysis of the modes was then carried out using a harmonic inversion code [21]. The boundary conditions in the *z* direction are perfectly matched layers so as to avoid any additional fold of the bands.

Based on those considerations, the in-plane band diagram of the structure is reported in Figure 3. It should be emphasized that a similar band diagram (not shown here) was obtained for a weakly absorbing layer ($\kappa \leq 10^{-2}$), which is the real case for a-Si at 720 nm. The black and red symbols are associated with the symmetric and anti-symmetric guided modes, respectively.

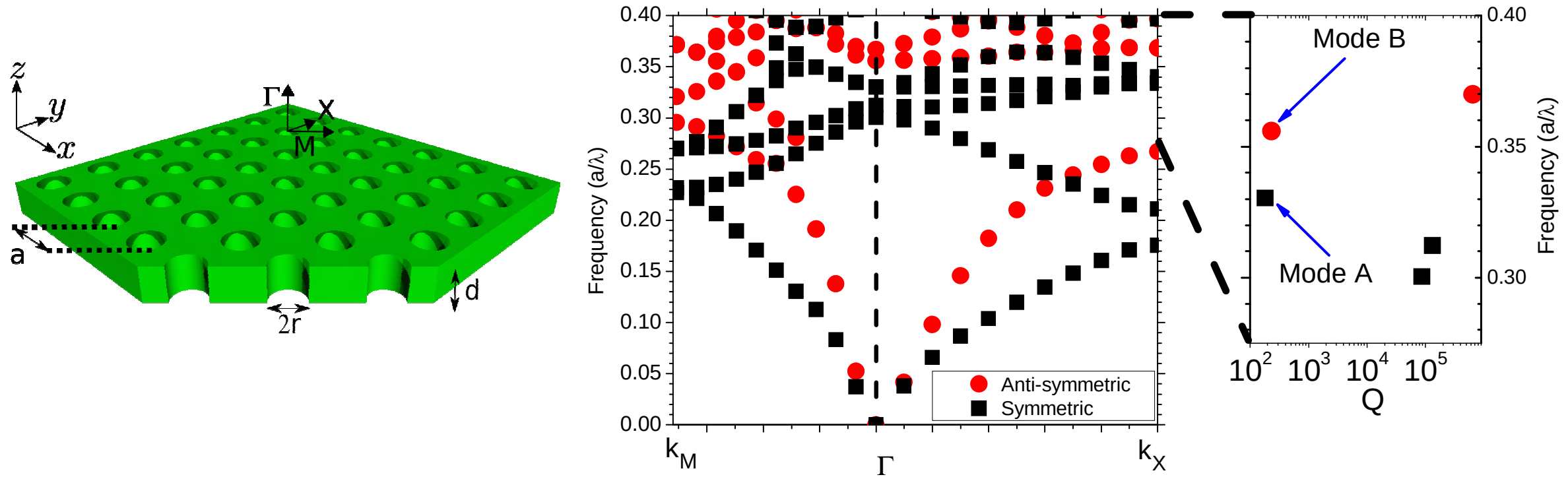

**FIGURE 3 : LEFT: SCHEMATIC VIEW OF THE 2D PHC MEMBRANE WITH ITS CORRESPONDING PARAMETERS. CENTER: BAND DIAGRAM OF THIS NON ABSORBING MEMBRANE WITH D=A/2 AND FF =0.15 AND FOR BOTH SYMMETRIC (OR EVEN) (BLACK SYMBOLS) AND ANTI-SYMMETRIC (OR ODD) (RED SYMBOLS) MODES. RIGHT: FREQUENCY VERSUS Q FACTOR OF THE MODES IN Γ ZOOMED IN THE FREQUENCY RANGE OF INTEREST.**

In the vertical direction, the two modes with the lowest frequency which can couple with the free space are denoted modes A and B on Figure 3. The other modes, exhibiting very large Q factors (not infinite because of numerical approximations), cannot couple to free space due to symmetry reason [22].

Figure 4 presents the mappings over one period and the $E_x$ field profile of both modes A and B across the membrane.

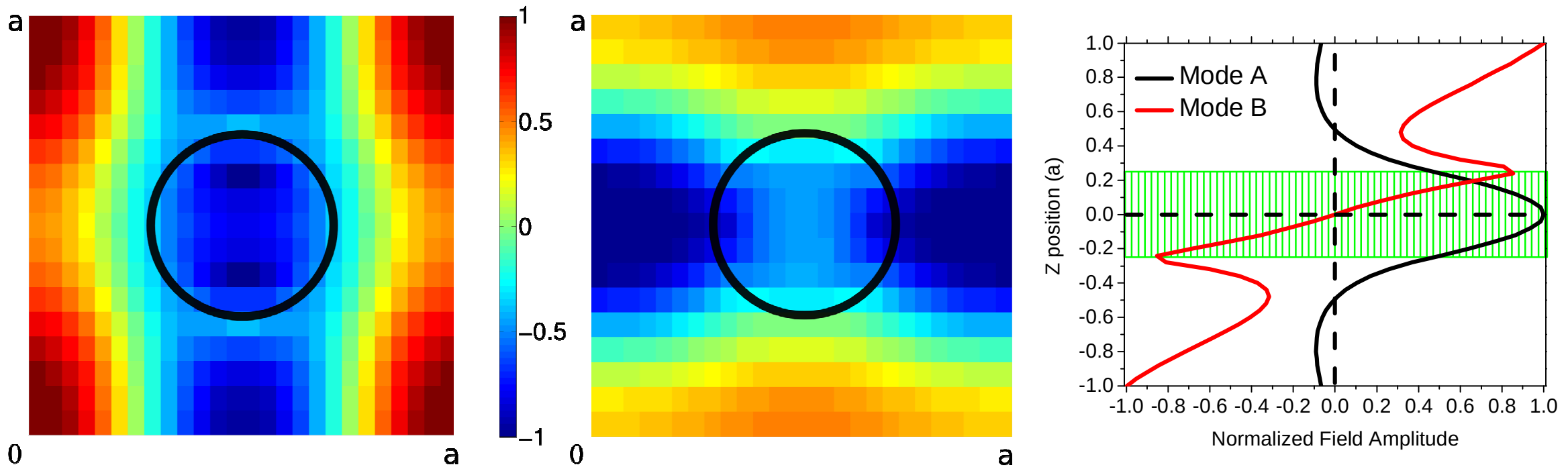

**FIGURE 4 : MAPPING OF MODE A (LEFT) AND MODE B (CENTER) OVER ONE PERIOD AND $E_X$ FIELD AMPLITUDE IN CROSS SECTION FOR THOSE TWO MODES (RIGHT. MODES A AND B ARE REPRESENTED BY A BLACK AND A RED LINE, RESPECTIVELY). ON THE MAPPINGS, THE HOLE IS DEPICTED BY A BLACK CIRCLE WHILE FOR THE CROSS SECTION VIEW, THE MEMBRANE CORRESPOND TO THE GREEN DASHED AREA. THE PARAMETERS OF THE MEMRBANE ARE FF=0.15, D=a/2 AND K=0.**

It can be checked that the considered modes are symmetric and anti-symmetric, as required by the analytical model. They are thus good candidates to satisfy the theoretical conditions mentioned in section 2 for maximizing the absorption in such a PhC membrane.

In the following, it is first checked that, in absorbing material these modes can be described using the TDCMT. Then, the influence of *ff* and *d* on their modal properties is detailed for a given $\kappa$. These parameters are adjusted to reach the required degeneracy for a maximized absorption. To get rid of the period dependency, all the parameters with a length dimension were normalized by the period *a*.

*TDMCT modelling of the selected PhC modes*

The $Q_{tot}$ and absorption (A=1-R-T) variations with $\kappa$ obtained by FDTD simulations are plotted on Figure 5.

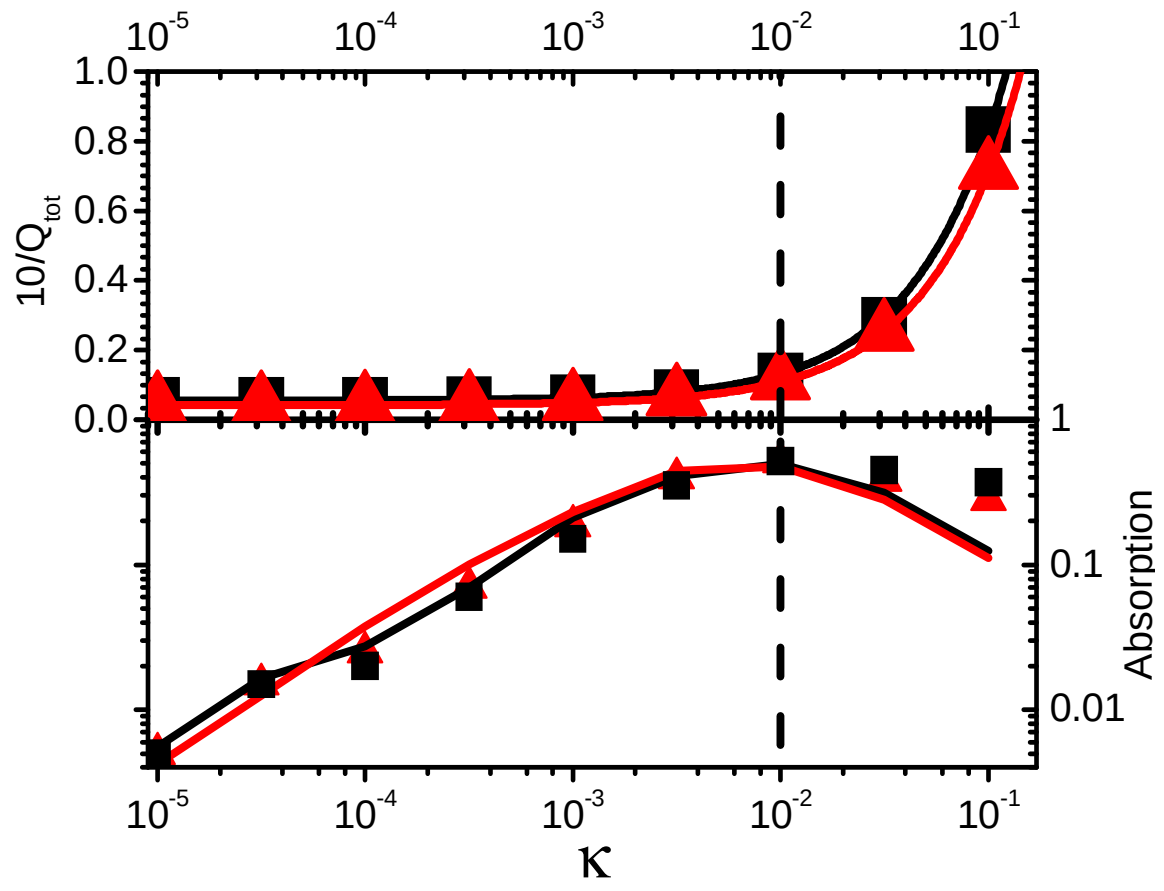

**FIGURE 5: 10 / $Q_{TOT}$ (TOP) AND ABSORPTION (BOTTOM) VARIATIONS VERSUS EXTINCTION COEFFICIENT FOR MODE A (IN BLACK) AND B (IN RED), D=A/2 AND ff=0.15. THE VALUES (SQUARE FOR MODE A, TRIANGLE FOR MODE B) REPORTED ARE FITTED (LINES) BY THE THEORETICAL MODELS DERIVED FROM (7) FOR Q AND FROM (8) FOR THE ABSORPTION. THE DASHED LINE REPRESENTS THE K VALUE SELECTED FOR THE REST OF THE STUDY (CRITICAL COUPLING CONDITIONS)**

With a linear scale in $\kappa$ (not shown here), the values of $10/Q_{tot}$ describe a straight line. This means that the losses are additives. Consequently, the data were fitted with the function described in (7).

$$\frac{1}{Q_{tot}(k)} = \gamma \times \kappa + \frac{1}{Q_e} \qquad (7)$$

where $Q_e$ $(= \omega_0 \tau_e)$ describes the coupling with the propagative wave, and where the intrinsic losses of the mode ($Q_0$) are described as the product of $\kappa$ by a coupling coefficient, γ, which can be related to the mode confinement in the absorbing medium, as discussed before.The parameters resulting from the fit of the two curves are summarized in Table I. To be noted that the $Q_e$ parameter is not a free one but is actually fixed by the simulations performed for the membrane without absorption. The fit well approximates the simulated data since the error ($1\text{-}R^2$) is really low.

**Table I : Fit parameters of data from Figure 5 (top) obtained by the function given in (7), *Qe* values are not free parameter but extracted from FDTD simulations.**

| | Mode A | Mode B |
|---|---|---|
| $\gamma$ | 0.77 | 0.67 |
| $Q_e$ | 179 | 236 |
| Error ($1\text{-}R^2$) | $6\ 10^{-4}$ | $6\ 10^{-5}$ |

According to Figure 5 (bottom), the FDTD computed absorption exhibits two regimes. The first one at low κ is a growing absorption with $\kappa$, when the external coupling is stronger than the absorption. The second one (for high κ) is governed by absorption losses. In between, at κ $=10^{-2}$, is located the critical regime for which the absorbing losses equal the coupling ones giving a maximum absorption of 50% for symmetric or anti-symmetric modes (as indicated in (3) and [11]).

The losses were successfully fitted with an additive law (7), which proves that the hypotheses made to use TDCMT (1) are consistent with the simulations.

For mode A (respectively B), $Q_{tot}$ = 80 (92) is appropriate to reach 50% absorption. The corresponding required $Q_e$ value of 179(236) is almost twice $Q_{tot}$ taking into account the numerical precision of our calculation.

Since $\kappa$ is usually fixed by the material, these results mean that ff=0.15 is an appropriate choice, as verified later. It can finally be noticed in both cases that $Q_0$ is of the order of the predicted required value

So as to fully check the validity of our model, we expressed the absorption as a function of the variables used in (7) injected in (2) at the resonance, which gives:

$$A = \frac{2Q_e}{\gamma k\left(Q_e + \frac{1}{\gamma \times \kappa}\right)^2} \qquad (8)$$

As can be checked on Figure 5 (bottom) this TDCMT derived model fits well the FDTD results. Notice that for high κ the model underestimates the FDTD calculated absorption. Indeed, single path absorption is discarded and more importantly, Fabry-Pérot absorption, which starts to be significant [6], is neither taken into account, as explained at the beginning of section III.

At this stage, both the symmetric and anti-symmetric modes exhibit an optimized absorption of 50%, but at different wavelengths.

# IV. PHC PARAMETERS REFINEMENT GUIDELINES FOR ABSORPTION ENHANCEMENT

*Influence of the filling factor on the Q of the PhC selected modes*

The filling factor of the PhC is changed to assess how this parameter is impacting the modal properties. Figure 6 displays the quality factor ($Q \equiv \tau\omega$, in black) and frequency (in red) variations with *ff* changes with a constant thickness *d= a/2*. The material is again supposed to be transparent. Indeed, the variations of the modes frequencies with $\kappa$ are much weaker than the one related to the quality factor and are therefore neglected.

Then, only the coupling properties of the PhC modes are studied ($Q = Q_e \equiv \tau_e\omega$), and $Q$ has thus no upper limit due to the absorption.

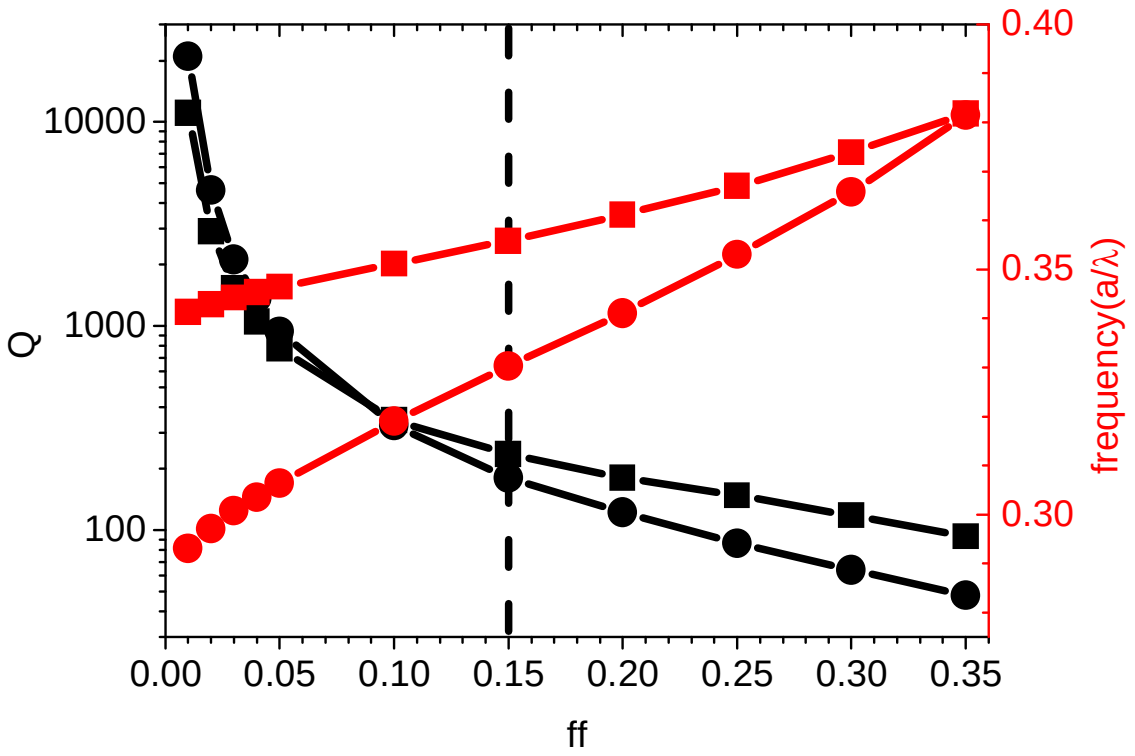

**FIGURE 6 : QUALITY FACTOR (IN BLACK) AND FREQUENCY (IN RED) OF THE MODES VERSUS FILLING FACTOR IN AIR FOR D=A/2. MODE A IS REPRESENTED BY SQUARES AND MODE B BY DOTS. THE DASHED LINE CORRESPONDS TO THE PARAMETERS CHOSEN FOR THE STUDY**

It can be noticed that as in [23] modifying *ff* allows to tune the quality factor of both modes (from about 50 to more than $10^4$ for reasonable values of *ff*) while keeping the frequency shift below 20%. Indeed, since both A and B modes are guided, their $Q$ is infinite without patterning (*ff*=0). Coupling is introduced by the PhC patterning, and light confinement decreases when ff increases, leading to larger coupling losses. The range of reachable $Q$ (=$Q_e$) is wide enough to address the critical coupling condition in the range of $0.002 \leq \kappa \leq 0.02$ using the previously described approximation (6).

Then, the filling factor appears to be the first parameter to be tailored so that the coupling constant with the plane wave enables the critical coupling conditions.

*Influence of the thickness on the frequency of the selected modes and spectral analysis of the absorbing membrane*

Lastly, the thickness of the membrane is adjusted to set the two modes at the same frequency, i.e to create a degeneracy, to maximise the absorption.

Figure 7 shows, for each mode, the transmission and reflection calculated using FDTD, and the resulting absorption, as well as the absorption modelled by (8), versus the thickness. In addition, the frequencies and $Q_{tot}$ of those modes are reported, together with the reflection, transmission and absorption spectra related to an absorbing membrane with a thickness corresponding to the absorption peak ($d$=0.68$a$) or away from this peak ($d$=0.5$a$).

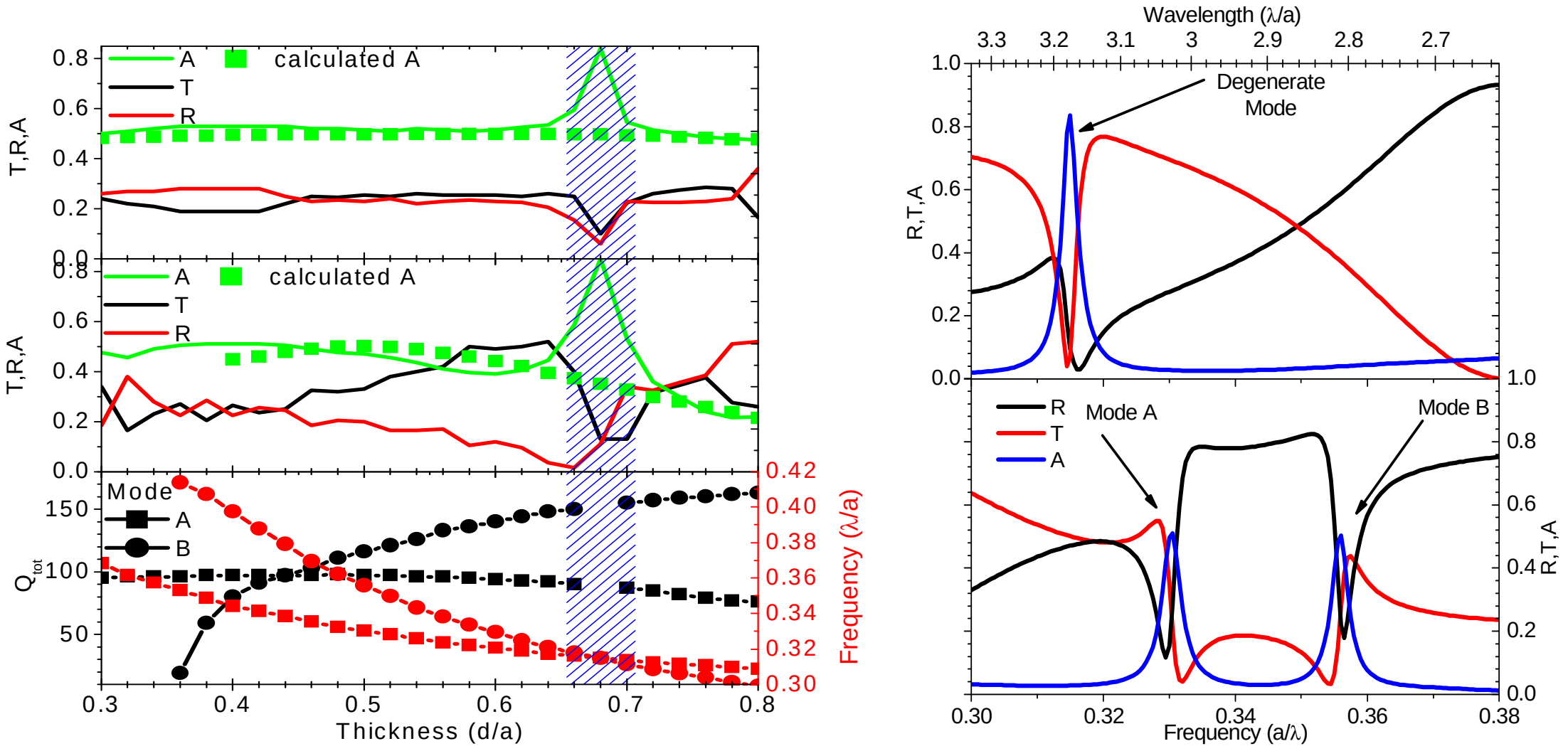

**FIGURE 7 : LEFT: TOP (MODE A) AND CENTER (MODE B): SIMULATED (FDTD) AND CALCULATED (TDCMT) (8) ABSORPTION (GREEN LINES AND GREEN SQUARES, RESPECTIVELY), TRANSMISSION (BLACK LINES) AND REFLECTION (RED) VERSUS MEMBRANE THICKNESS. BOTTOM: VARIATION OF THE MODAL PROPERTIES (Q IN BLACK AND FREQUENCY IN RED) WITH THE MEMBRANE THICKNESS FOR MODE A (SQUARES) AND MODE B (DOTS). RIGHT: ABSORPTION (BLUE), TRANSMISSION (RED) AND REFLECTION (BLACK) VERSUS FREQUENCY OF THE ABSORBING MEMBRANE WITH $\kappa=10^{-2}$. THE SPECTRA ON TOP CORRESPOND TO THE CREATION OF THE DEGENERATE STATE (D=0.68a) AND THE SPECTRA ON BOTTOM TO SEPARATED MODES A AND B (D=0.5a).**

On this figure, both simulated and modeled absorption of both modes are monotone functions of the thickness of the membrane, except for the simulated absorption at a single peak standing at $d$=0.68$a$.

More precisely, apart from the absorption peak, the absorption is limited to about 50% for both modes (when taking into account the numerical precision of our simulations). However, two different behaviors for modes A and B can be observed: While the absorption for mode A is almost constant at around 50%, substantial changes of the absorption occur for mode B. This can be explained by the variation of the $Q_{tot}$ of the modes, which is almost negligible for mode A, but is varying from 10 to 150 for mode B for the thicknesses studied. The variation of the vertical confinement with the thickness, thus of the coupling, might be stronger for mode B than for mode A. Indeed the field in mode A is more concentrated in the center of the membrane than in mode B. The variation of the absorption with $Q_{tot}$ can be directly related to the absorption variations as shown by the modeled absorption (8) curve.

For $d$=0.68$a$, the curves describing the frequency variations are crossing as can be seen on the bottom part of the left hand side of Figure 7. This gives rise to a degenerate state between modes A and B. This is made possible since the modal dispersion curves are not decreasing at the same rate when the membrane thickness is increased.

This particular state can be described by the TDCMT as exposed in (4). This degeneracy leads to an absorption peak overcoming the usual 50% absorption limit for the isolated PhC membrane. Its value is the sum of the calculated absorptions for modes A and B.

The spectral properties shown on Figure 7 (right part) confirm the origin of the absorption peak. Indeed, When $d$=0.5$a$, two absorption peaks are observed, corresponding to the two non degenerate modes, whereas the reflection and transmission exhibit a asymmetric Fano resonance shape typical of the coupling between the free space waves and the PC modes A Lorentzian fit associated to their absorption peaks gives a $Q$ of 97 and 116, respectively, close to the value obtained using Harminv. Conversely, when $d$=0.68$a$, those two modes are degenerate leading to an absorption of 85% with a peak width corresponding to a quality factor of 109, using the full width at half maximum of the peak.

To go further, one can compute the respective weight of modes A and B in the degenerate state, using (4) and introducing:

$$\frac{1}{\tau_{totA}} = \frac{1}{\tau_{eA}} + \frac{1}{\tau_{0A}} \tag{9}$$

With an equivalent expression for mode B, those coefficients can be written:

$$\alpha = \frac{\kappa_A \tau_{totA}}{\sqrt{\kappa_B{}^2 \tau_{totB}{}^2 + \kappa_A{}^2 \tau_{totA}{}^2}}$$
$$\beta = \frac{\kappa_B \tau_{totB}}{\sqrt{\kappa_B{}^2 \tau_{totB}{}^2 + \kappa_A{}^2 \tau_{totA}{}^2}} \tag{10}$$

Where α and β are the respective weight of modes A and B (in amplitude). It is then possible to calculate the modes weight in energy (using the decay rates values extracted from FDTD simulations) as:

$$\alpha^2 = 57\%$$
$$\beta^2 = 43\% \tag{11}$$

This result clearly shows the important contribution of both modes in the absorption. The latter reaches 85% for the degenerate mode, which is significantly higher than 50%, although noticeably lower than 100%.

Besides, a numerical refinement of the PhC parameters could further improve the absorption; and the robustness of the absorption with respect to technology and precision and experimental conditions could be appreciate through the first derivatives of corresponding curves, but is out of the scope of this paper.

In addition, so as to apply these results to broadband absorber, and since the TDCMT gives the width of the absorption peak via λ/Q, the spectral density magnitude of the peaks needed around one targeted wavelength λ can be deduced directly from the κ value of the material (7). As a matter of fact, it is roughly the $Q_{tot}$ value that can be approximated by $\kappa^{-1}$. In other words, given a Δλ range around λ, the order of magnitude of modes spectral density needed to optimize the absorption at 50% is about $\kappa^{-1}$Δλ /λ.

# V. CONCLUSION

To summarize, we first introduced a model based on the TDCMT to describe the absorption within an isolated absorbing layer through resonant mode, showing that the coupling anisotropy of the mode is one of the key points to reach total absorption, well beyond the usual 50% admitted limit. This model allows us to predict the intensity and the spectral width of the absorption peak taking place at one given wavelength with only few modal parameters, such as the quality factor of the modes considered. Subsequently, using FDTD simulations, we varied the 2D photonic crystal thin membrane parameters in order to reach high (85%) absorption using degenerate modes while the absorption in an equivalent unpatterned layer would be really low (since $\kappa$=$10^{-2}$). To achieve this, we derived a methodology for tailoring the absorption which can be extended to other systems involving a thin photonic absorbing membrane.

One highlight of this study was to raise the importance of the anisotropy in the system. Actually, the benefit from anisotropy has already been exploited through the integration of a back reflector [12, 6]. This may spectrally impact the absorption due to interference effects differently than for anisotropic patterns such as the ones presented in [16, 17, 18]. Using these anisotropic patterns for tailoring the absorption provides additional degrees of freedom which can even be combined to a back reflector.

Finally, this method can be directly applied to indoor photovoltaic or sensor applications for which the absorption must be well-controlled but does not need a broad spectral range. In addition, it can also be extended to solar photovoltaic using a collection of modes as presented in [24, 10]. One way to achieve this would then be to work in a higher frequency domain on the band diagram of the patterned membrane so as to take profit from a higher density of modes.

# VI. ACKNOWLEDGEMENT

R.P. would like to thank OrangeLabs and G.G. the Rhône-Alpes region for the financial support of this study.